\documentstyle[a4,12pt]{article}

\input epsf

\begin {document}
\newcommand{\etal}{{\it et al.}}
\title
{ 
Interacting Agents and Continuous Opinions Dynamics
}
\author{
 G{\'e}rard~Weisbuch$^{*}$ \\
Guillaume Deffuant$^{**}$, Frederic Amblard$^{**}$ \\
and Jean Pierre Nadal$^{*}$
\\
$^{*}$Laboratoire de Physique Statistique\footnote{
Laboratoire associ{\'e} au CNRS (URA 1306), {\`a} l'ENS et
 aux Universit{\'e}s Paris 6 et Paris 7}
\\
     de l'Ecole Normale Sup{\'e}rieure, \\      
    24 rue Lhomond, F-75231 Paris Cedex 5, France. \\
$^{**}$Laboratoire d'Ing{\'e}nierie pour les Syst{\`e}mes Complexes (LISC)\\
Cemagref - Grpt de Clermont-Ferrand\\
24 Av. des Landais - BP50085\\
F-63172 Aubi{\`e}re Cedex (FRANCE)\\
  {\small     {\em email}:weisbuch@lps.ens.fr}\\ 
 }
\maketitle

\abstract{We present a model of opinion dynamics in which
agents adjust continuous opinions as a result of
random binary encounters whenever their  difference in opinion
 is below a given threshold. High thresholds
yield convergence of opinions towards an average opinion, 
whereas low thresholds result in several opinion clusters.
The model is further generalised to network interactions,
threshold heterogeneity, adaptive thresholds and binary
strings of opinions.}

\section{Introduction}
  Many models about opinion dynamics (F\"olmer 1974, Arthur 1994, 
Orl\'ean 1995,
 Latan\'e and Nowak 1997, Weisbuch and Boudjema 1999),
are based on binary opinions
which social actors update as a result of social influence,
often according to some version of a majority rule.
  Binary opinion dynamics
have been well studied, such as the herd behaviour described 
 by economists (F\"olmer 1974, Arthur 1994, Orl\'ean 1995).
One expect that in most cases
the attractor of the dynamics will display uniformity of
opinions, either 0 or 1, when interactions occur across the whole
population. 
 Clusters of opposite opinions
 appear when the 
dynamics occur on a social network with exchanges restricted
to connected agents. Clustering is reinforced
when agent diversity, such as a disparity in influence,
 is introduced, (Latan\'e and Nowak 1997, Weisbuch and Boudjema 1999).

 One issue
of interest concerns the importance of the binary assumption:
what would happen if opinion were a continuous variable such as 
the worthiness of a choice (a utility in economics), or some belief
about the adjustment of a control parameter? 

  The rationale for binary versus continuous opinions
might be related to the kind of information used by agents  
to validate their own choice:
\begin{itemize}
\item the actual choice of the other agents, 
a situation common in economic choice of brands:
"do as the others do"; 
\item the actual opinion of the other agents,
about the "value" of a choice: "establish one's opinion
 according to what the others think or at least according
to what they say".
\end{itemize}

On the empirical side, there exist
 well documented studies
about social norms concerning sharing between partners.
 Henrich {\it et al.} (2001) compared through experiments
shares accepted in the ultimatum game and showed 
that people agree upon what a "fair" share should be,
which can of course vary across different cultures.
Young and Burke (2000) report empirical data 
about crop sharing contracts, whereby a landlord leases his farm to a tenant
laborer in return for a fixed share of the crops. 
In Illinois as well as in India, crop sharing distributions
are strongly peaked upon "simple values" such as 1/2-1/2
or 1/3-2/3. The clustering of opinions about "fair shares"
is the kind of stylised fact that our model tries to reproduce.
More generally, we expect such opinion dynamics to occur
in situation where agents have to make important choices
and care to collect many opinions before taking any decisions:
adopting a technological change might often be the case.
The present paper was motivated by changes towards
environmental-friendly practices in agriculture
under the influence of the new Common Agricultural Policy    
 in Europe.

 Modeling
of continuous opinions dynamics
was earlier started by applied mathematicians
and focused on the conditions under which
a panel of experts would reach a consensus, (Stone 1961, Chatterjee
and Seneta 1977, Cohen \etal 1986, Laslier 1989, Krause 2000,
Hegselmann and Krause 2002, further referred to as the "consensus"
literature).
 
The purpose of this paper
is to present results concerning continuous opinion dynamics
subject to the constraint that convergent
 opinion adjustment only proceeds
when opinion difference is below a given threshold.
The rationale for the threshold condition is that agents only 
interact when their opinions are already close enough; 
otherwise they do not even bother to discuss. The reason
for refusing discussion might be for instance lack of understanding, 
conflicting interest, or social pressure. The threshold would then 
correspond to some openness character.
Another interpretation is that the threshold corresponds to
uncertainty: the agents have some initial views with some degree
of uncertainty and would not care about other
views outside their uncertainty range. 

 Many variants of the basic idea can be proposed
and the paper is organised as follows:
\begin{itemize}
\item 
We first expose the simple case of complete mixing 
among agents under a unique and constant threshold
condition.
\item 
We then check the genericity of the results obtained
for the simplest model to other cases such as
localised interactions, distribution of thresholds, varying thresholds and 
binary strings of opinions.
\end{itemize}

  A previous paper (Deffuant \etal 2000) reported more complete results
on mixing across a social network and
on binary strings of opinions.

\section{The basic case: Complete Mixing and one fixed threshold}
  
  Let us consider a population of $N$ agents $i$
with continuous opinion $x_i$. We start from an initial distribution 
of opinions, most often taken uniform on [0,1] in the computer simulations.
 At each time step any two 
randomly chosen agents
meet:
 they re-adjust their opinion when their difference in 
 opinion is smaller in magnitude than a threshold $d$.
Suppose that the two agents have opinion $x$ and $x'$.
$Iff$ $|x-x'|<d$ opinions are adjusted according to:
\begin{eqnarray}
  x &=& x + \mu \cdot (x'-x) \\
  x' &=& x' + \mu \cdot (x-x') 
\label{eq1}
\end{eqnarray}

 where $\mu$ is the convergence parameter
whose values may range from 0 to 0.5.

In the basic model, the threshold $d$ is taken as
 constant in time and across the whole population. 
Note that we here apply a complete mixing hypothesis
plus a random serial iteration mode\footnote{The "consensus" 
literature most often uses parallel iteration mode when they
suppose that agents average at each time step the opinions
of their neighbourhood. Their implicit rationale
 for parallel iteration is that they model successive
 meetings among experts. 
}.

The evolution of opinions may be mathematically predicted
in the limiting case of small values of $d$ (Neau 2000)\footnote{The
other extreme is the absence of any threshold which yields
consensus at infinite time as earlier studied in Stone 1961 and others.}.
 Time variations of opinions' density $\frac{d \rho(x)}{dt}$ 
obey the following dynamics: 
$$
\frac{d \rho(x)}{dt}= - \frac{d^3}{2}\cdot \mu \cdot (1-\mu) \cdot
\frac{\partial ^2(\rho ^2)}{{\partial x}^2}
$$
 This implies that
 starting from an initial distribution of opinions
in the population, any local higher opinion density is amplified.
Peaks of opinions increase and valleys are depleted
until very narrow peaks remain among
 a desert of intermediate opinions.

  For finite thresholds,
computer simulations show that the distribution of opinions
evolves at large times towards clusters of homogeneous opinions.

\begin{figure}[!h]
\centerline{\epsfxsize=90mm\epsfbox{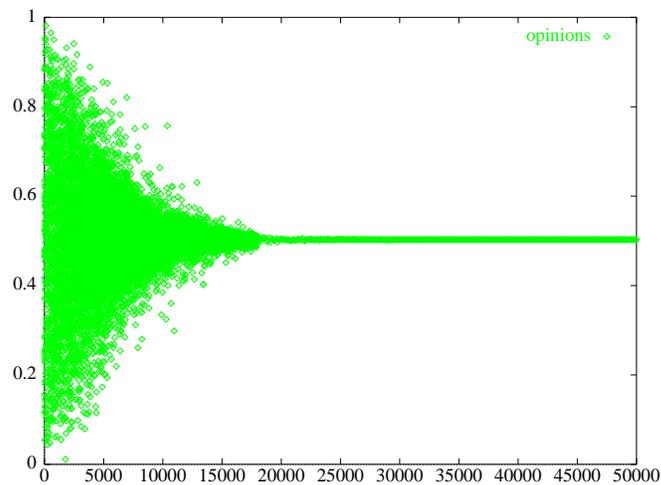}}
\label{fig:convergence}
\caption{Time chart of opinions ($d=0.5 \quad \mu=0.5 \quad N=2000$).
One time unit corresponds to sampling a pair of agents.} 
\end{figure}
\begin{itemize}
\item 
 For
large threshold values ($d>0.3$) only one cluster
is observed at the average initial opinion.
Figure 1 represents the time evolution of opinions
starting from a uniform distribution of opinions.
\item 
For lower threshold values, several clusters can be observed
(see figure 2). Consensus is then NOT achieved when thresholds are low
enough. 
\end{itemize}
\begin{figure}[!ht]
\centerline{\epsfxsize=90mm\epsfbox{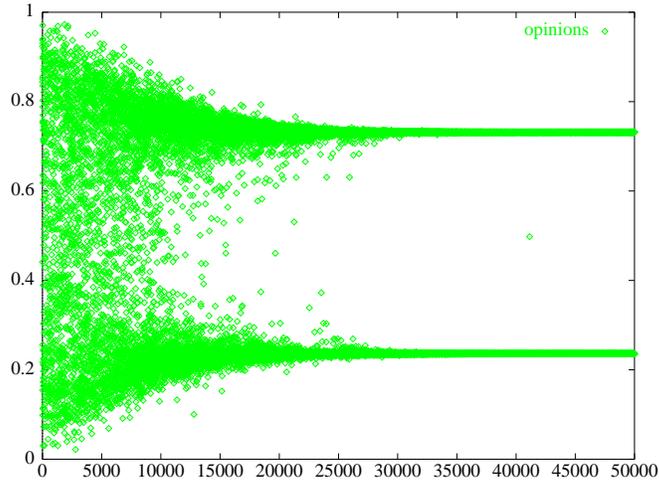}}
\label{2pics}
\caption{Time chart of opinions for a lower threshold $d=0.2$ ($ \mu=0.5, \quad N=1000$)
.} 
\end{figure}
Obtaining clusters of different opinions does not
surprise an observer of human societies, but this result
was not a priori obvious since we started from an initial
configuration where transitivity of opinion
propagation was possible through the entire population: 
any two agents however different in opinions
could have been related through a chain of agents with closer 
opinions. The dynamics that we describe ended up in
gathering opinions in clusters on the one hand, but also in
separating the clusters in such a way that agents in different
clusters don't exchange anymore.   

The number of clusters varies as the integer part of $1/2d$:
this is to be further referred to as the "1/2d rule"
(see figure 3\footnote{Notice the continuous transitions
in the average number of clusters when $d$ varies. 
Because of the randomness of the initial distribution and pair
sampling, any prediction on the outcome of dynamics
such as the 1/2d rule can be expressed as true with a probability close
to one in the limit of large $N$; but one can often generate a deterministic
sequence of updates which would contradict the "most likely" prediction.}).
\begin{figure}[!ht] 
\centerline{\epsfxsize=90mm\epsfbox{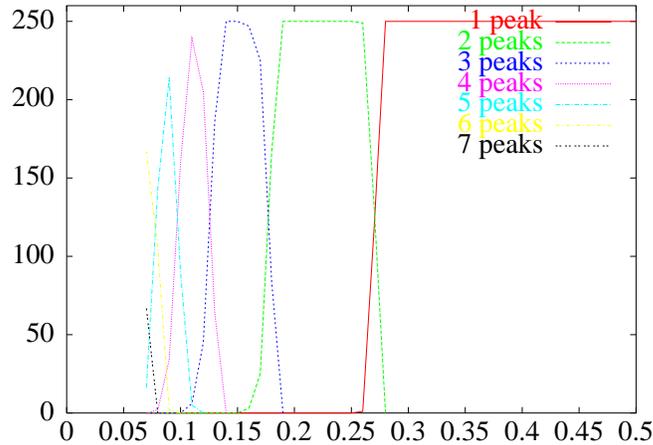}}
\caption{\label{fig:phases} Statistics of the number of opinion clusters
as a function of $d$ on the x axis for 250 samples ($\mu=0.5, \quad
N=1000$).} 
\end{figure}

\section{Social Networks}

  The literature on social influence and social choice also
considers the case when interactions occur along social connections 
between agents (F\"olmer 1974)
 rather than randomly across the whole population.
 Apart from the similarity condition, we now
 add to our model a condition on proximity, i.e.
 agents only interact if they are directly connected through a
pre-existing social 
relation. Although one might certainly consider
the possibility that opinions on certain un-significant subjects
could be influenced by complete strangers, we expect important decisions
to be influenced by advice taken either from professionals (doctors, for instance)
or from socially connected persons (e.g. through family, business, or clubs).
Facing the difficulty of inventing a credible instance of a social
network as in the literature on social binary choice, we here
adopted the 
standard simplification and carried out our simulations on square lattices:
square lattices are simple, allow easy visualisation of opinion
configurations and contain many short loops, a property that they 
 share with real social networks.

  We then started from a 2 dimensional network of connected
agents on a square grid.  Any agent can only interact with his four
connected neighbours (N, S, E and W). 
We used the same initial random sampling of opinions from 0 to 1 and
the same basic interaction process between
agents as in the previous sections. At each time step
a pair is randomly selected among {\it connected agents}
and opinions are updated according to equations 1 and 2
provided of course that their distance is less than $d$.

 The results are not very different from those observed with 
non-local opinion mixing as described in the previous section,
 at least for the larger values of $d$ ($d>0.3$, all results
displayed in this section
are equilibrium results at large times).
 As seen in figure 4, the lattice
is filled with a large majority of agents who have reached consensus
around $x=0.5$ while a few isolated agents have ``extremists''
opinions closer to 0 or 1. The importance of extremists is the 
most noticeable difference with the full mixing case described in the
previous section.

\begin{figure}[htbp]
\centerline{\epsfxsize=60mm\epsfbox{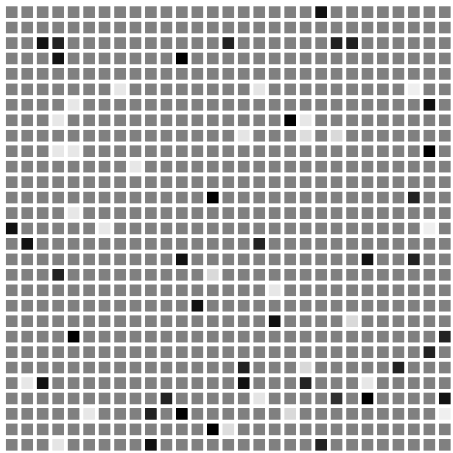}}
\label{fig:ru}
\caption{Display of final opinions of agents connected on a square
  lattice of size 29x29 ($d=0.3 \quad \mu=0.3$ after 100 000
  iterations) .
Opinions between 0 and 1 are coded by gray level (0 is black and 1 is white).
Note the percolation of the large cluster of homogeneous opinion
and the presence of isolated ``extremists''.} 
\end{figure}

  Interesting differences are noticeable for the smaller
values of $d<0.3$ as observed in figure 5.

\begin{figure}[htbp]
\centerline{\epsfxsize=60mm\epsfbox{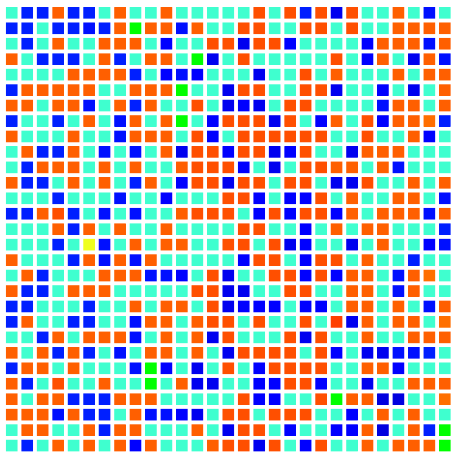}}
\label{fig:ru}
\caption{Display of final opinions of agents connected on a square
  lattice of size 29x29 ($d=0.15 \quad \mu=0.3$ after 100 000 iterations) .
Color code: purple 0.14, light blue 0.42, red 0.81 to 0.87.
Note the presence of smaller clusters with similar
but not identical opinions.} 
\end{figure}
 For connectivity 4 on a square lattice, only cluster
 percolates (Stauffer and Aharony 1994) across the lattice.
All agents of the percolating cluster share the same opinion
as observed on figure 4. But for  $d<0.3$
several opinion clusters are observed and none percolates
across the lattice. Similar opinions, but not identical, 
are shared across several clusters. 
 The differences of opinions between
groups of clusters relate to $d$, but the actual values
inside a group of clusters fluctuate from cluster to
cluster because
homogenisation occurred independently among the 
different clusters: the resulting opinion depends on fluctuations of
initial opinions and histories from one cluster to the other. 
The same increase in fluctuations compared to the full mixing case
is observed from sample to sample with the same parameter values.

  The qualitative results obtained with 2D lattices
should be observed with any connectivity, either periodic,
random, or small world.

 The above results where obtained when all agents have the same
invariant threshold. The purpose of the following sections
is to check the general character of our conclusions:
\begin{itemize}
\item when one introduces a distribution of thresholds in the population;
\item when the thresholds themselves obey some dynamics.
\end{itemize}

\section{Heterogeneous constant threshold}
\label{sec:het}
  Supposing that all agents use the same
threshold to decide whether to take into account
the views of other agents is a simplifying assumption.
When heterogeneity of thresholds is introduced,
some new features appear. To simplify the
matter, let us exemplify the issue in the case of a bimodal distribution of
thresholds, for instance 8 agents with a large threshold of 0.4
and 192 with a narrow threshold of 0.2 as in figure 6.    

\begin{figure}[htbp]
\centerline{\epsfxsize=120mm\epsfbox{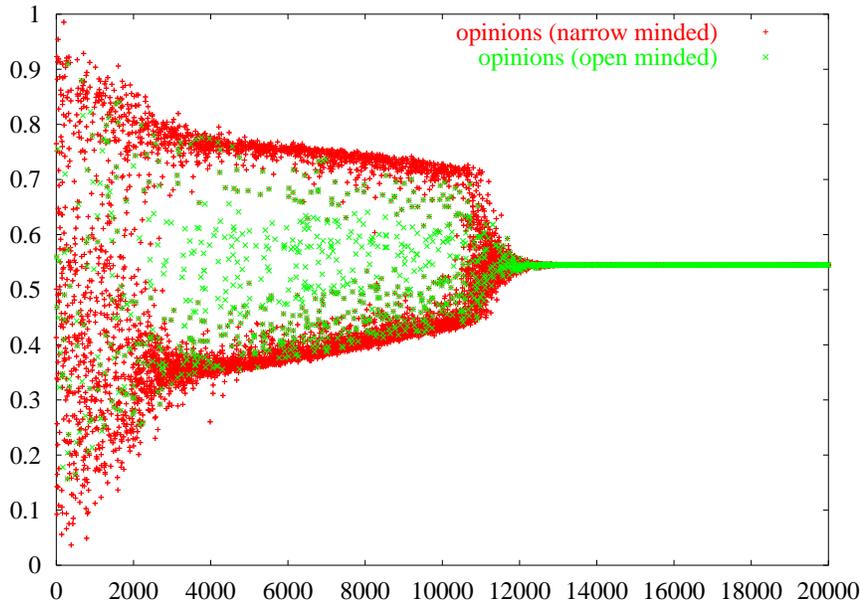}}
\label{conv}
\caption{Time chart of opinions ($ N=200$).
Red '+' represent narrow minded opinions (192 agents with threshold 0.2),
green 'x'   represent open minded opinions (8 agents with threshold 0.4).
} 
\end{figure}

 One observes that in the long run
 convergence of opinions into 
one single cluster is achieved due to the presence of 
the few "open minded" agents  (the single cluster convergence time
is 12000, corresponding to 60 iterations per agent on average,
 for the parameters of figure 6). But in the short run,
a metastable situation with two large opinion clusters close
to opinions 0.35 and 0.75 is observed due to narrow minded agents,
with open minded agents opinions fluctuating around 0.5
due to interactions with narrow minded agents belonging to either
high or low opinion cluster. 
Because of the few exchanges with the high $d$ agents,
low $d$ agents opinions slowly shift towards the average until 
the difference in opinions between the two clusters
falls below the low threshold: at this point the two
clusters collapse.

This behaviour is generic for any mixtures
of thresholds. At any time scale, the number of clusters obeys
a "generalized 1/2d rule":
\begin{itemize}
\item on the long run clustering depends on the higher threshold;
 \item on the short run clustering depends on the lower threshold;
\item the transition time between the two dynamics
is proportional to the total number of agents
and to the ratio of narrow minded to open minded agents. 
\end{itemize}

In some sense, the existence of a few "open minded" agents seems
sufficient to ensure consensus after a large enough 
time for convergence. The next section restrict the validity of this prediction
when threshold dynamics are themselves taken into account.

\section{Threshold Dynamics}
  
 Let us interprete the basic threshold rule 
in terms of agent's uncertainty: agents take into account others' opinion
on the occasion of interaction because they are not certain about
 the worthiness of a choice. They engage in interaction only with
 those agents which opinion does not differ too much from their own opinion
in proportion of their own uncertainty.
 If we interprete the threshold for exchange
as the agent uncertainty, we might suppose 
with some rationale that his subjective uncertainty
 decreases with the number of opinion exchanges.

Taking opinions from other agents can be
 interpreted, at least by the agent himself,
as sampling a distribution of opinions.
As a result of this sampling,
 agents should update their
new opinion by averaging over their 
previous opinion and the sampled external opinion
and update the variance of the opinion distribution 
accordingly.

Within this interpretation, a "rational procedure"
(in the sense of Herbert Simon) for the agent
is to simultaneously update his opinion
and his subjective uncertainty.
Let us write opinion updating as
 weighting one's  previous opinion
 $x(t-1)$ by $\alpha$ and the other 
agent's opinion $x'(t-1)$ by  $1-\alpha$, with $0<\alpha<1$.
$\alpha$ can be re-written $\alpha=1-\frac{1}{n}$
where $n$ expresses a characteristic number of opinions
taken into account in the averaging process.
$n-1$ is then a relative weight of the agent previous
opinion as compared to the newly sampled opinion
weighted $1$.
Within this interpretation, 
updates of both opinion $x$ and variance $v$ should be written:
\begin{eqnarray}
  x(t) &=& \alpha \cdot x(t-1) + (1-\alpha) \cdot x'(t-1) \\
  v(t) &=& \alpha \cdot v(t-1) +  \alpha (1-\alpha) \cdot [x(t-1) - x'(t-1)]^2 
\end{eqnarray}
 The second equation simply represents the change in variance
when the number of samples increases from $n-1$
at time $t-1$ to $n$ at time $t$.
It is directly obtained from the definition
of variance as a weighted sum of squared deviations.

As previously, updating occurs when the difference in opinion 
is lesser than a threshold, but this threshold is now related
to the variance of the distribution of opinions
sampled by the agent.
  A simple choice is to relate the threshold to the
standard deviation $\sigma(t)$ according to:
  \begin{equation}
    d(t)= \nu \sigma(t) ,
  \end{equation}
where $\nu$ is a constant parameter often taken equal to 1
in the simulations.

When an agent equally values collected opinions 
independently of how old they are, he should also update $\alpha$
connected to $n-1$ the number of previously collected opinions:
\begin{equation}
  \alpha(t) = \frac{n(t)-1}{n(t)} \alpha(t-1)
\end{equation}
 This expression is also used in the literature about
"consensus" building to describe "hardening" 
of agents opinions as in Chatterjee and Seneta (1977) and Cohen \etal
in 1986.

Another possible updating choice is to maintain
 $\alpha$ constant which corresponds to taking
a moving average on opinions and giving more importance to
the $n$ later collected opinions. Such a "bounded" memory would
make sense in the case when the agent believes that there exists
some slow shift in the distribution of opinions, whatever
its cause, and that older opinions should be discarded. 

 Both algorithms
were tried in the simulations and give qualitatively similar results
in terms of the number of attractors, 
provided that one starts from an initial number of supposed trials $n(0)$
corresponding to the same $\alpha$.

\subsection{Scaling}
\subsubsection{Constant  $\alpha$}
In the case of constant $\alpha$, a simplified
computation valid in the limit of small $\nu$ predicts 
an exponential decay of thresholds.
Neglecting the second term in the dynamics of variance 
\footnote{In fact adding the second term would compensate the decay in variance due to 
the multiplication by $\alpha$ in the limit of large $\nu$;
for finite $\nu$, partial compensation depends on the form of
the distribution of opinions, but anyway,
variance decays exponentially with a smaller rate than when $\nu$
is close to 0} :

\begin{equation}
   v(t) = \alpha \cdot v(t-1)
\end{equation}
gives :
\begin{equation}
   v(t) = \alpha^t \cdot v_0 \\ .
\end{equation}

Writing $\alpha$ as $1-\frac{1}{n}$ and approximating 
it as exp$(\frac{-1}{n})$ for large $n$,
we see that the variance decays exponentially with
a characteristic time of $n$ and that the thresholds
vary as:
 \begin{equation}
   d(t) \simeq d_0  \cdot exp [- \frac{t}{2n}]
\end{equation}
 A parallel estimation for the dynamics of convergence of
opinions towards some average opinion $x_{\infty}$ (corresponding to the attractor)
can be made by replacing $x'(t-1)$ by $x_{\infty}$ in 
 equation (3) describing the dynamics
of $x(t)$. After subtracting $x_{\infty}$ to both members,
the deviation of opinions from their attractor 
can be written as: 
\begin{equation}
   x(t)-x_{\infty} = \alpha \cdot (x(t-1)-x_{\infty})
\end{equation}
 Equation 10 shows that opinions also converge exponentially towards the attractor with the
same time constant as variance.

\subsubsection{Varying  $\alpha$}

 Equivalent computations were also made
for the case when $ \alpha$ varies as $1- \frac{1}{n(t)}$.
For instance, the dynamics of variance is described by the following
set of equations:
\begin{equation}
   v(t) =  (1- \frac{1}{n(t)}) \cdot v(t-1)
\end{equation}
\begin{equation}
   v(t) = v_{0} \prod_{n'=n_0}^{n_0+t} (1- \frac{1}{n'})
\end{equation}
and with some approximation:
\begin{equation}
   v(t) \simeq v_0 \frac{n_0}{n_0+t}
\end{equation}
  Thresholds then vary as the inverse square root of the number of 
interactions.

 The equivalent computation for the evolution of opinion deviation
from the attractor also gives an hyperbolic decay:
\begin{equation}
   x(t)-x_0 \simeq (x(0)-x_{\infty}) \frac{ n_0}{n_0+t}
\end{equation}

The above expressions allow us to predict average trends 
for the dynamics of thresholds and opinions.
\begin{itemize}
\item The variable $t$ appearing in the expressions
is NOT time, but the number of ACTUAL updates of the
agent opinions. The time scaling laws, exponential or hyperbolic are the same
 for both variance and distance to the attractor.
When going from time to number of updates,
one should take into account the frequency of sampling one agent
and the probability of actually updating the agent which is proportional to
the inverse number of attractors. With this caveat, the simulations results
are in accordance with the above predictions.
\item The scaling laws are different for constant and varying $\alpha$,
with faster convergence (exponential) when $\alpha$ is kept constant.
But one should note that in both cases opinions dynamics follow
the same scaling rules as thresholds dynamics: 
phenomena such as clustering should then be similar.
 The same dynamics of opinions and thresholds are to
be observed in both conditions provided that the horizontal axis used to
plot opinions for varying $\alpha$ is warped to an exponential for constant 
 $\alpha$.
\end{itemize}

\subsection{Simulation Results}

 When comparing to constant threshold dynamics,
decreasing thresholds results in a larger variety of 
final opinions. For initial thresholds values 
which would have ended in opinion consensus,
one observes a number of final clusters which decreases with $\alpha$
(and thus with $n$). Smaller values of $\alpha$ correspond
to a fast decrease of the thresholds, which prevents
 the aggregation of all opinions into large clusters.

Observing the chart of final opinions versus initial opinions
on figure 7,
one sees that most opinions converge towards two clusters
(at $x=0.60$ and $x=0.42$)
which are closer than those one could obtain with constant thresholds
(typically around $x=0.66$ and $x=0.33$):
initial convergence gathered opinions which would had aggregate 
at the initial threshold values, but which later segregate due to
the decrease in thresholds. Many outliers are apparent on the plot.   

\begin{figure}[htbp]
\centerline{\epsfxsize=120mm\epsfbox{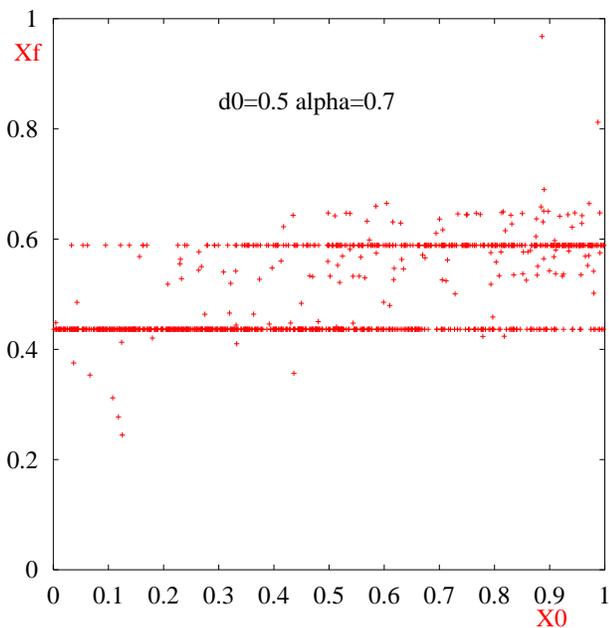}}
\label{fig:2pics}
\caption{Each point on this chart represents the final opinion
of one agent versus its initial opinion 
 (for constant $\alpha=0.7 \quad \nu=1.0 \quad N=1000$,
initial threshold 0.5).
 } 
\end{figure}

\subsubsection{Large  $\alpha$ and $n$}

 Large values of $\alpha$, close to one, e.g. $n>7$, correspond to averaging on 
many interactions.
The interpretation of large $\alpha$ and $n$ is that the
agent has more confidence in his own opinion than in the opinion of the
other agent with whom he is interacting, in proportion with $n-1$.

For constant values of $\alpha$, the observed dynamics
is not very different
from what we obtained with constant thresholds.
 
 The exponential decay of thresholds predicted by equation 9
is verified on figure 8 plotted for the same parameters values. 
The observed decay constant on figure 8 is 1.7, slightly less
than 2, the theoretical prediction based on equation 9 which 
 neglects the possible increase of variance
due to other opinions.
  
\begin{figure}[htbp]
\centerline{\epsfxsize=100mm\epsfbox{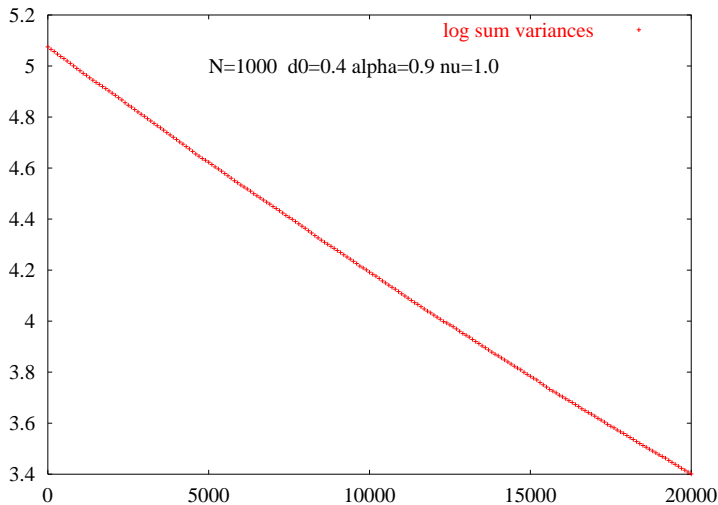}}
\caption{Exponential decay of summed variances
(for constant $\alpha=0.9 \quad  d(0)= 0.4 \quad \nu=1.0 \quad N=1000$,
initial threshold 0.4).}
\end{figure}

For  $ \alpha$ 
varying according to equation 13,
 the variance dynamics is  
hyperbolic as observed on the log-log
plot of figure 9. The observed slope on figure
9 is not far from the predicted value, -1.0.

 In both cases the scaling of variance and thresholds is verified 
on more than one decade, but deviations are observed:
\begin{itemize}
\item for varying  $\alpha$, clustering is slow and deviations
are observed at small times when the segregation of clusters 
is not yet achieved because the probability of updating a
random pair is not yet constant;
\item for constant  $\alpha$, deviations
are observed at large times due to the existence of outliers
which maintain a larger variance (see the next section). 
\end{itemize}

\begin{figure}[htbp]
\centerline{\epsfxsize=100mm\epsfbox{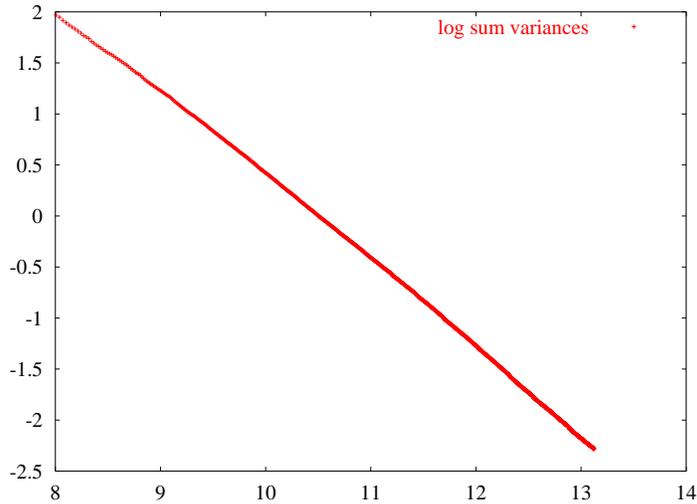}}
\caption{Power-law decay of summed variances
for varying $ \alpha$
 ( log-log plot, initial $\alpha=0.7 \quad d(0)= 0.4 \quad \nu=0.5 \quad N=200$).}
\end{figure}

\subsubsection{Small  $\alpha$ and $n$}

 \begin{figure}[htbp]
\centerline{\epsfxsize=120mm\epsfbox{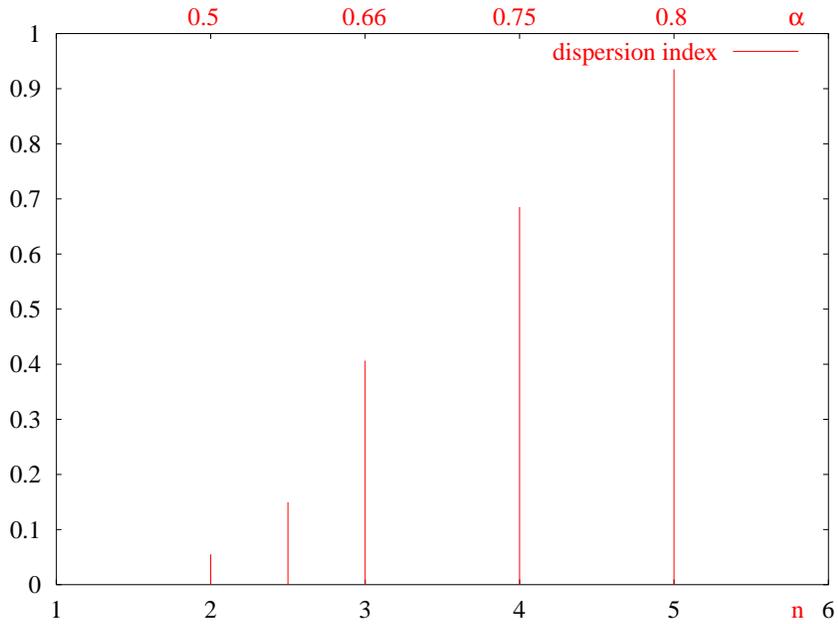}}
\label{fig:disp}
\caption{ Variation of the dispersion index $y$ with $n$,
 the initial "subjective" number of collected opinions 
( $\alpha=1-1/n, \quad  d(0)= 0.5 \quad \nu=1.0 \quad N=1000$).
The initial threshold value of 0.5 if kept constant would yield consensus 
with only one cluster.} 
\end{figure}

  A more complicated dynamics is observed for lower values of 
$n$ and $\alpha$. Apart from the expected main clusters,
one also observes large and
small clusters plus isolated individuals (outliers).

 For $d(0)=0.5$ (which would yield consensus 
with only one cluster if kept constant) and $\alpha=0.5$ (corresponding to
$n=2$, i.e. agents giving equal weight to their
own opinion and to the external opinion),
 more than ten clusters unequal in size are observed
plus isolated outliers. One way to characterise the dispersion
of opinions with varying $\alpha$ is to compute $y$ the relative value of
the squared cluster sizes with respect to the squared number of opinions.

\begin{equation}
  \label{eq:xx}
  y = \frac{\sum_{i=1}^n s_i^2}{(\sum_{i=1}^n s_i)^2}
\end{equation}

 For $m$ clusters of equal size, one would have $y=1/m$.
 The smaller $y$, 
the more important is the dispersion in opinions.
Figure 10 shows the increase of the dispersion index $y$ with $n$
($n-1$ is the initial "subjective" weight of agent's own opinion).

As previously noticed on figure 7, some outliers
do not aggregate in the large clusters.
 The origin of these isolated agents is due to randomness of:
\begin{itemize}
\item sampling the individual agents at various times;
\item sampling the pairs, i.e. which pair of agent is sampled
for possible interaction.
\end{itemize}

\begin{figure}[htbp]
\centerline{\epsfxsize=120mm\epsfbox{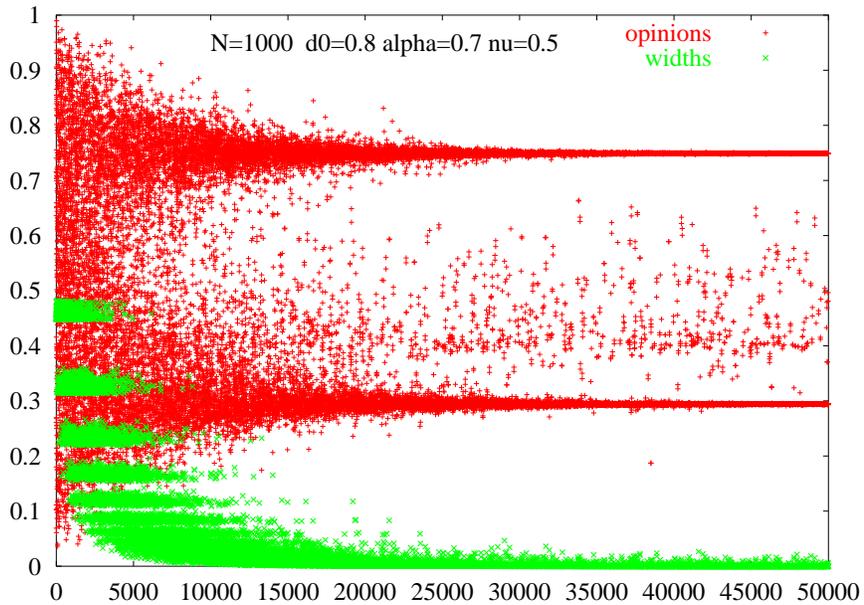}}
\label{fig:2pics}
\caption{Time chart of opinions 
and thresholds (for constant $\alpha=0.7 \quad  d(0)= 0.4 \quad \nu=0.5 \quad N=1000$).
 Red '+' represent opinions and 
green 'x'   represent thresholds.} 
\end{figure}

The time pattern of thresholds appearing as green 
bands on figure 11 give us some insight on these effects.
 One band corresponds to a given number of opinion exchanges
experienced by the agents:
the upper band corresponds to the variance after one exchange, the second upper
to two exchanges and so on. The lower bound of a band corresponds to 
the result of interactions between very close opinions when the second term
in equation 4 is negligible. The vertical width of a band
is due to this second term, which relative importance to
the first can be estimated from the figure: it 
is roughly 10 perc. (for $\nu=0.5$).
The horizontal width of a band corresponds to the fact that 
different agents are experiencing the same number of updating
 at different times:
 rough evaluations made on figure 11 show that most agents have their
first exchange between time 0 and 4000, and their fifth exchange between
1000 and 12 000.  

When the decrease of threshold and the clustering 
of opinions is fast, those agents which are 
not sampled early enough and/or not paired with close 
enough agents can be left over from the clustering process.
 When they are sampled later, they might be too far 
from the other agents to get involved into opinion
adjustment. The effect gets important when convergence is fast,
i.e. when $n$ and $\alpha$ are small.

 Let us
note that these agents in the minority have larger uncertainty 
and are more "open to discussion" than those in the mainstream,
in contrast with the common view that eccentrics
are opinionated!
 For $\alpha=0.7$, $d(0)= 0.4$, $\nu=0.5$ and $N=1000$,
the parameters of figure 11, we found that mainstream agents 
in the two attractors account for 43 and 42 perc. of the population
while 15 perc. are in the minority peaks.
 
  The results of the dynamics are even more dispersed for
lower values of $\alpha$.
 In this regime, corresponding to "insecure agents"
who don't value their own opinion more than those
of other agents, we observe more clusters which
importance and localisation depend on the random sampling
of interacting agents and are thus harder to predict 
than in the other regime with a small number of big clusters. 
Using a physical metaphor, clustering in this regime resembles
quenching to a frozen configuration, thus maintaining many "defects"
(e.g. here the outliers), while the opposite
regime it resembles annealing (with suppression of defects). 

  In fact, one way to evaluate the influence of the two effects,
randomness of the time at which agents are sampled
from randomness of pairing, is to compare the standard random 
updating iteration that we have used to parallel updating:
in a parallel updating, a random pair matching of
all agents is first realised and all pairs are then updated 
simultaneously.  Parallel updating then suppresses
randomness of updating time: only randomness of pairing remains.

We found by comparing the two algorithms,
random and parallel, for the same set of parameters, that
parallel updating only slightly reduces the number of outliers.
We can then conclude that 
 most of the observed disorder results from
the randomness of pairing. 

\subsection{Distribution and Dynamics of thresholds }
  Finally, an obvious set of simulations to perform
is to have a distribution of thresholds and to let
these thresholds evolve according to one of the two rules,
 $\alpha$ constant or decreasing, according to equation 5.

\begin{figure}[htbp]
\centerline{\epsfxsize=120mm\epsfbox{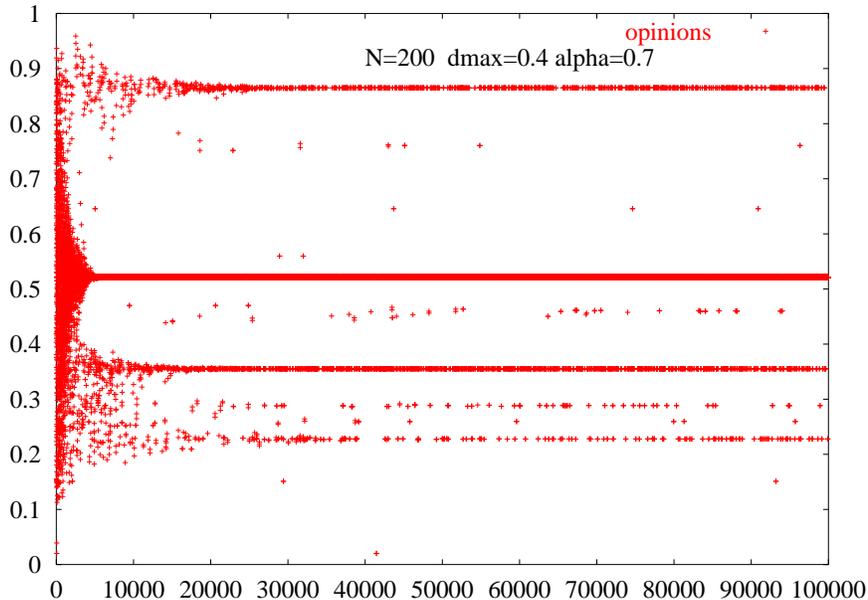}}
\label{fig:2pics}
\caption{Time chart of opinions for
a distribution of varying thresholds
(constant $\alpha=0.7  \quad \nu=1 \quad N=200$).
The initial thresholds are uniformly distributed on $[0,0.4]$.
} 
\end{figure}

We used a uniform distribution of thresholds on $[0, d_{max}]$
 and observed, not surprisingly since some thresholds are 
close to zero even at initial times, that clusters and outliers
are in larger number than for single initial thresholds.
 Figure 12 displays the time evolution
 of opinions for a constant 
value of $\alpha=0.7$ ,  $\nu=1$ and  $N=200$.
The initial thresholds were uniformly distributed on $[0,0.63]$.
Clusters correspond to agents with larger thresholds, outliers
to thresholds close to 0. 

 Convergence times differ according
to the size of the clusters: agents in large clusters have 
more occasions to update their opinion in proportion
to the number of agents in the same cluster. Small clusters then
need longer times to get stabilised. 

These results generalise and summarise our previous findings
in sections 4, 5.1 and 5.2.

\section{Vector opinions}
\label{sec:vv}
\subsection{The model}
  Another subject for investigation is vectors of opinions.
Usually people have opinions on different subjects, which can be
represented by vectors of opinions. In accordance with our
previous hypotheses, we suppose that one agent
interacts concerning different subjects with another agent according to
some distance with the other agent's vector of opinions.
In order to simplify the model, we revert to binary opinions.
An agent is characterised by a vector of $m$ binary opinions
about the complete set of $m$ subjects, shared by all agents. 
We use the notion of Hamming distance between binary opinion vectors
(the Hamming distance between two binary opinion vectors
is the number of different bits between the two vectors).
Here, we only treat the case of complete mixing; any pair of agents
might interact and adjust opinions according to how many opinions  
they share.\footnote{The bit string model shares some resemblance with
Axelrod's model of disseminating culture (Axelrod 1997)
based on adjustment of cultures as sets of vectors of integer variables
characterising agents on a square lattice.}
 The adjustment process occurs when agents agree
on at least $m-d$ subjects (i. e. they disagree
on $d-1$ or fewer subjects).
The rules for adjustment are as follows: 
 when opinions on a subject differ, one agent 
(randomly selected from the pair) is
convinced by the other agent with probability $\mu$.
Obviously this model has connections with
population genetics in the presence of sexual
recombination when reproduction only occurs if genome distance is smaller 
than a given threshold. Such a dynamics results in the emergence of species
 (see Higgs and Derrida 1991). 
We are again interested in the clustering of opinion vectors.
In fact clusters of opinions here play
 the same role as biological species in evolution. 

\subsection{Results}

We observed once again that $\mu$ and $N$ only modify convergence
times towards equilibrium;  the most influential factors
are threshold $d$ and $m$ the number of subjects under discussion.
Most simulations were done for $m=13$. For $N=1000$,
convergence times are of the order of 10 million pair iterations.
For $m=13$:
\begin{itemize}
\item When $d>7$, the radius of the hypercube, convergence
towards a single opinion occurs (the radius of the hypercube
is half its diameter which is equal to 13, the maximum distance in
the hypercube).
\item Between $d=7$ and  $d=4$ a similar convergence is observed for
more than 99.5 per cent of the agents with the exception
  of a few clustered or isolated opinions distant from
the main peak by roughly 7.
\item For $d=3$, one observes from 2 to 7 significant peaks
(with a population larger than 1 per cent) plus some isolated opinions.
\item For $d=2$ a large number (around 500) of small clusters is
  observed
(The number of opinions is still smaller than the maximum number
of opinions within a distance of 2).  
\end{itemize}

 The same kind of results are obtained with larger values of $m$:
two regimes, uniformity of opinions for larger $d$ values and extreme diversity
for smaller $d$ values, are separated by one $d_c$ value
for which a small number of clusters is observed (e.g for $m=21$,
$d_c=5$. $d_c$ seems to scale in proportion with  $m$
). 

Figure 13 represents these populations 
of the different clusters at equilibrium
(iteration time was 12 000 000) in a log-log plot according to
their rank-order of size. No scaling law is obvious from these plots,
but we observe the strong qualitative difference
in decay rates for various thresholds $d$.

\begin{figure}[htbp]
\centerline{\epsfxsize=120mm\epsfbox{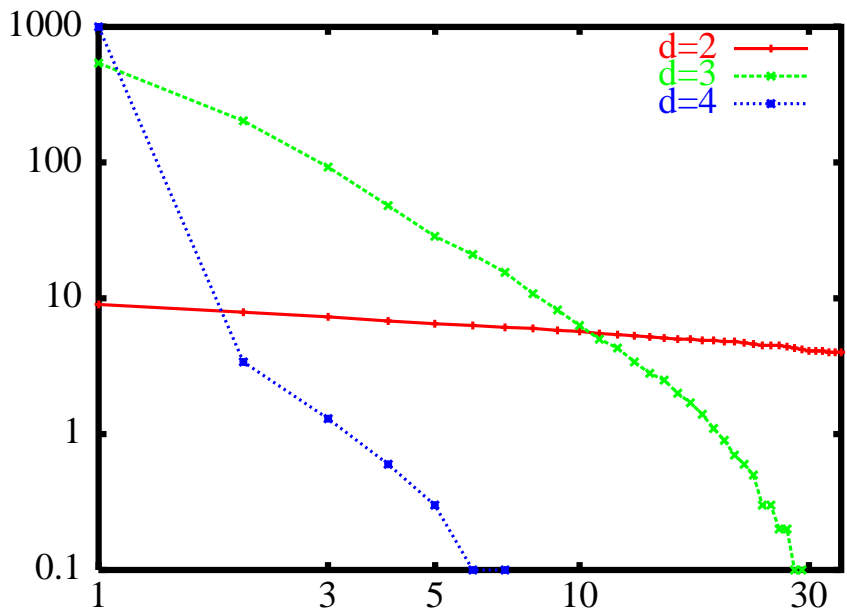}} 
\label{fig:2pics}
\caption{Log-log plot of average populations of clusters of opinions
arranged by decreasing order
for $N=1000$ agents ($\mu=1$).} 
\end{figure}

\section{Conclusion}

  The main lesson from this set of simulations is 
that opinion exchanges restricted by a proximity
threshold result into clustering of opinions. The $1/2d$ rule
predicts the outcome of the dynamics in the simplest 
cases, but it also provides some qualitative insight
for the case of threshold dynamics.

  Continuous opinions and binary strings share some
similarity: clustering with a number of clusters
decreasing with $d$. But continuous opinions
display a regular decrease of cluster number 
with $d$ while  binary strings display a phase
transition from consensus to a large multiplicity of clusters.

    When one introduces dynamics on thresholds
on the basis that agents interprete opinion
exchange as sampling a distribution of opinions:
\begin{itemize}
\item The amplitude of $\alpha$ and $n$
determine clustering properties.
 \item For large $\alpha$ and $n$, when agents trust their
own opinion more that the opinion of any other agent,
 updating is smooth resulting in large clusters (annealing).   
\item For small $\alpha$ and $n$, with relatively "insecure" agents,
updating is irregular resulting
in more clusters and outliers (quenching).
 \item 
Maintaining $\alpha$ constant, (short term memory)
or updating also $\alpha$ each time a new opinion is collected
does not change the outcome of the clustering process
but changes the convergence time: constant $\alpha$
yields a fast exponential convergence, while varying $\alpha$
results in an hyperbolic convergence. 
\end{itemize}

 One can of course question the genericity of the 
results that were obtained with such simple models.
In fact the main result, namely clustering, would not
be canceled but rather re-inforced by two most direct
generalisations of the model: 
\begin{itemize}
\item A given population of agents might have
a distribution of eventually conflicting interests
which could be translated in our formalism as initial
clustering of opinions or at least a non-uniform distribution.
\item
Opinions can also result from the combination of hypotheses which can already 
lead to different clusters because of conflicting interpretations.
\end{itemize}

We can then conclude that clustering into different opinions groups is the 
rule as soon as openness is limited.

  \bigskip

{\bf Acknowledgments}:

 We thank David Neau
and the members of the IMAGES FAIR project,
  Edmund Chattoe, Nils Ferrand and Nigel Gilbert
  for helpful discussions. GW benefited at different stages
in the project from discussions with Sam Bowles, Winslow Farell
and John Padgett at the Santa Fe Institute whom we 
thank for its hospitality. Thanks to Rainer Hegselmann
for pointing us the references to the "consensus" literature.
This study
 has been carried out with financial support
    from the Commission of the European Communities, Agriculture and Fisheries
    (FAIR) Specific RTD program, CT96-2092, "Improving Agri-Environmental
    Policies : A Simulation Approach to the Role of the Cognitive Properties of
    Farmers and Institutions". It does not necessarily
    reflect its views and in no way
    anticipates the Commission's future policy in this area. 

\bigskip
{\bf References}

Arthur, B. W. (1994)
``Increasing Returns and Path Dependence in the Economy'',
 University of Michigan Press, Ann Arbor, MI.       

Axelrod R. (1997) "Disseminating cultures" in Axelrod R.,
The complexity of cooperation, Princeton University Press. 

Chatterjee S. and Seneta E., (1977)
"Towards consensus: some convergence theorems on
repeated averaging",  {\em J. Appl. Prob.} 14, 89-97.

Cohen J. E., Hajnal J. and Newman C.M. (1986)
"Approaching consensus can be delicate when 
positions harden"  {\em Stochastic Processes and their Applications}
22, 315-322.

G. Deffuant, D. Neau, F. Amblard and G. Weisbuch (2000)
"Mixing beliefs among interacting agents"
 {\em Advances in Complex Systems} 3, 87-98. 

F\"ollmer H. (1974) "Random Economies with Many Interacting Agents",
 {\em Journal of Mathematical Economics}
1/1, 51-62.

Hegselmann R. and Krause U. (2002) "Opinion formation 
under bounded confidence" proceedings of the Simsoc5 conference
 to appear in JASSS.

Henrich J., Boyd R., Bowles S., Camerer C.,
Fehr E., Gintis H., and McElreath R. (2001)
          "In Search of Homo Economicus: Behavioral Experiments in 15
          Small-Scale Societies"
    {\em  Am. Econ. Rev.} 91, 2, pp.  73-78

 Higgs P.G. and Derrida, B. (1991)
``Stochastic models for species formation in evolving populations'',
 {\em J. Phys. A: Math. Gen.} 24, 985-991.

Krause U. (2000) "A discrete non-linear and non-autonomous model
of consensus formation" in Communications in Difference Equations,
Elaydi etal edit. Gordon and Breach pub.

Laslier, J.F. (1989)
``Diffusion d'information et {\'e}valuations s{\'e}quentielles''
 {\em Economie appliqu\'ee}.

Latan\'e, B. and Nowak, A. (1997) "Self- Organizing Social Systems:
 Necessary and Sufficient Conditions for the
Emergence of Clustering, Consolidation and Continuing Diversity",
pp. 43-74  in Barnett, G. A.
 and Boster, F. J. (eds.) Progress
in Communication Sciences.       

Neau, D (2000), ``R\'evisions des croyances dans 
un syst\`eme d'agents en interaction'', rapport d'option
de l'\'ecole polytechnique, available at 
http://www.lps.ens.fr/\~{ }weisbuch/rapneau.ps.

Orl\'ean A. (1995), "Bayesian Interactions and Collective Dynamics of
Opinions: Herd Behavior and Mimetic Contagion", {\em Journal of
Economic Behavior and Organization}, 28, 257-274. 

Stauffer D. and Aharony A. (1994) "Introduction to Percolation Theory",
 Taylor and Francis, London.

Stone M. "The opinion Pool" (1961)
 {\em Ann. of Math. Stat.} 32, 1339-1342.

Weisbuch G. and Boudjema G. (1999), ``Dynamical aspects in the
Adoption of Agri-Environmental Measures'',  {\em Adv. Complex Systems} 2,
11-36.     

  Young H. P., and Burke M. A. (2001)
 "Competition and Custom in Economic Contracts: A Case Study
 of Illinois Agriculture", {\em  Am. Econ. Rev.},
 91, 3, pp. 559-573.

\end {document}